\documentclass{aipproc}

\usepackage{graphics}
\usepackage{color}
\usepackage{here}
\usepackage{epsfig}

\usepackage{longtable}
\def\to{\rightarrow}

\def\ra{\rightarrow}
\newcommand\nutau{{\nu_\tau}}

\newcommand\numu{{\nu_\mu}}

\newcommand\nue{{\nu_e}}

\def\dm2{\Delta m^2}
\def\sq2{sin^2(2\Theta)}

\def\NOE{{\em N\raise.5ex\hbox{O \kern-0.47em}E\kern.4em}}

\def\ICANOE{{\em ICANOE}}
\def\FLUKA{{\sc FLUKA}}

\def\Journal#1#2#3#4{{#1} {\bf #2}, #3 (#4)}
\def\etal{{\it et\ al.}}

\def\PLB{{\em Phys. Lett.}  B}

\begin{document}

\title{ICANOE \\ Imaging and Calorimetric Neutrino Oscillation Experiment}
\author{Andr\'e Rubbia$^1$\footnote{On behalf of the ICARUS and NOE
Collaborations. Talk given at the Workshop on the Next generation Nucleon decay and Neutrino detector (NNN99), September 23-25, 1999.}}
\affiliation{\it $~^1$ Institut f\"{u}r Teilchenphysik, ETHZ, CH-8093 Z\"{u}rich, Switzerland}

\begin{abstract}
The main scientific goal of the \ICANOE{} detector\cite{icanoe}
is {\it the one of elucidating in a comprehensive way 
the pattern of neutrino masses and mixings}, following the SuperKamiokande
results and the observed solar neutrinos deficit.
To achieve these goals, the experimental method is based upon the 
{\it complementary and simultaneous detection of CERN beam (CNGS) and 
cosmic ray (CR)  events}.  
For the currently allowed values 
of the SuperKamiokande results, both CNGS and cosmic ray data will give independent 
measurements and provide a precise 
determination of the oscillation
parameters.  
Since one will observe and unambiguously identify
$\nue$, $\numu$ and $\nutau$ components, 
the full (3 x 3) mixing matrix will be explored.
\end{abstract}

\maketitle

\section{Introduction}

The reference mass for underground
detectors is now set by the operating SuperKamiokande\cite{superk} detector, 
which is of the order of 30 ktons.  However the rather coarse nature of the Cherenkov 
ring detection is capable to reconstruct only part 
of the features of the events. 

New generation underground experiments are now facing new challenges, 
for which novel and 
more powerful technologies are required, with respect to the existing detectors:  
\begin{enumerate}

\item The {\it long baseline accelerator neutrino oscillation} experiments require, with 
respect to existing short baseline detectors (like NOMAD\cite{nomad} and CHORUS\cite{chorus}), a
large 
increase of the 
detector fiducial mass (about 3 ton in NOMAD), in order to cope with the 
flux attenuation due to the distance. In order to perform a 
comprehensive program on neutrino oscillations the fiducial mass of  
the detector must be increased to several ktons.
Moreover, the detector must be able to tag efficiently 
the interaction of $\nue$'s and $\nutau$'s out of the bulk
of $\numu$ events. This requires a detailed event reconstruction
that can be achieved only by means of a high granularity
detector. 

\item Likewise, the comprehensive investigation of 
{\it atmospheric neutrinos} events, in order to reach the level of at 
least one thousand 
events/year, also requires a fiducial mass of several ktons. 
The capability to observe all separate processes, 
electron, muon and tau neutrino charged
currents (CC) and all neutral currents (NC) 
without detector biases and 
down to kinematical threshold is highly desirable.

\item {\it Nucleon decay}: because of the already very high limit on 
the nucleon lifetime ($\ge 
10^{32}$ years in most of the decay channels), a modern proton decay detector 
should have an adequately large sensitive mass.

\end{enumerate}

Event imaging should be provided by a modern bubble 
chamber-like technology since
(1) it has to be able to provide high resolution, unbiased, three dimensional 
images of ionising events; 
(2) it has to provide an accurate measurement of the basic kinematical 
properties of the particles of the event, including particle identification. 
(3) it has to accomplish simultaneously the two basic functions of target and 
detector.

In the \ICANOE{} design a fully
sensitive, bubble chamber-like detector will 
permit discovery limits at the few events level and a much more powerful background 
rejection. A detector of this
kind, already at the level of a few ktons of mass will be 
fully competitive with the potentialities of SuperKamiokande and in several domains will 
permit to extend much further the investigations.  

The \ICANOE{} detector fruitfully merges the superior
imaging quality of the ICARUS technology\cite{icaruswww} with the high resolution full 
calorimetric containment of NOE\cite{noewww}, suitably upgraded to provide also
magnetic analysis of muons. It has a modular structure 
of independent supermodules 
and is expandable by the addition of such supermodules,
each consisting
of a low density 1.9 kton liquid target and of a high density 
0.8 kton active solid target.

The superior quality of the event vertex inspection and 
reconstruction of the liquid Argon is ideally 
complemented by the addition of the external module 
capable of magnetic analysis of the muons escaping the 
LAr chamber. Bubble chambers have in fact often been 
very similarly complemented in the past by external identifiers. 
An iron muon tracking spectrometer would fulfill this job,
but it would also introduce in between adjacent liquid argon
volumes a blind region incapable of giving information on
the energy and on the nature of the  
escaping particles. A sensitive magnetized calorimeter appears 
therefore as an ideal containment module to be interleaved between
adjacent liquid Argon volumes.

\section{Outline of the \ICANOE{} detector}

\begin{figure}[tb]
\centering
\epsfig{file=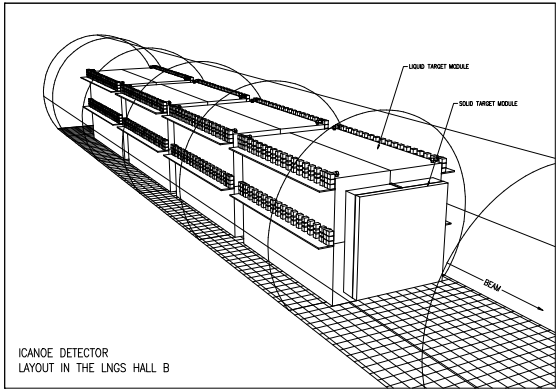,width=5.5cm,angle=-90}
\caption{Perspective view of the baseline detector with
4 supermodules.}
\label{icanoe_pr}
\end{figure}

The \ICANOE{} layout (Figure~\ref{icanoe_pr})
is similar to that of a ``classical'' neutrino detector,
segmented into almost independent {\bf supermodules}. 
The layout of the apparatus can be summarized as follows:
\begin{itemize}
\item the {\bf liquid target}, with extremely high resolution,
dedicated to tracking, $dE/dx$ measurements, 
full e.m. calorimetry and hadronic calorimetry, where
electrons and photons are identified and measured with extremely
good precision and $\pi/\mu$, $K$ and $p$ separation is possible
at low momenta;
\item the {\bf solid target}, with good e.m. and hadronic resolution,
dedicated to calorimetry of the jet
and a magnetic field for measurement of the muon features
(sign and momentum);
\item The {\bf supermodule}, obtained joining a liquid and solid module,
which constitutes the basic module of an {\it expandable} apparatus.
A supermodule 
behaves as a complete building block, capable of identifying and measuring electrons,
photons, muons and hadrons produced in the events. The solid, high density sector 
reduces the transverse and longitudinal size of hadron shower, confining the event
(apart from the muon) within the supermodule. 
\end{itemize}

The \ICANOE{} {\it SuperModule} is, according to the present design, 
composed by a liquid Argon module, with 18.0$\times$ 
(11.3)$^2$ m$^3$ of external dimensions and 1.4~kton (1.9~kton) of
active (total) mass, and a magnetized calorimeter module, 
with 2.6$\times$ 9$^2$ m$^3$ of external dimensions 
and 0.8~kton of mass. The depth consists of  
 2~m for the calorimetric unit, corresponding 
to $7.4\ \lambda_{int}$ and $59\ X_0$ and $0.6$~m for the tracking units 
(the interleaved planes of tracking chambers).

At this stage, {\it four} SuperModules with 
a total length of the experiment of $82.5$~m and a total active mass 
of 9.3~kton fully instrumented are being considered
for the baseline option.

\section{Physics goals}
\ICANOE{} is 
an underground detector capable to achieve the full reconstruction of 
neutrino (and antineutrino) events of {\it any} flavor, and with an
energy ranging from the tens of MeV to the tens of GeV, for the relevant
physics analyses. No other combinations can provide such a rich spectrum
of physical observations, 
including the systematic, on-line monitoring of the CNGS $\nu$-beam at the 
LNGS site.
The unique lepton capabilities of \ICANOE{} are really fundamental in
tagging the neutrino flavor. In general, the oscillation pattern 
of the neutrinos may be complicated
and involve a combination of $\numu\ra\nutau$, $\numu\ra\nue$ and
$\numu\ra\nu_{sterile}$ transitions. In order to fully sort out
the mixing matrix, unambiguous neutrino flavor identification is
mandatory to distinguish $\tau$'s from $\nu_\tau$'s and electrons
from $\nu_e$'s interactions. In other words, we stress the importance
of constraining the oscillation scenarios by coupling appearance in
several different channels 
and disappearance signatures.

  \begin{figure}[tb]
    \centering
\mbox{\epsfig{figure=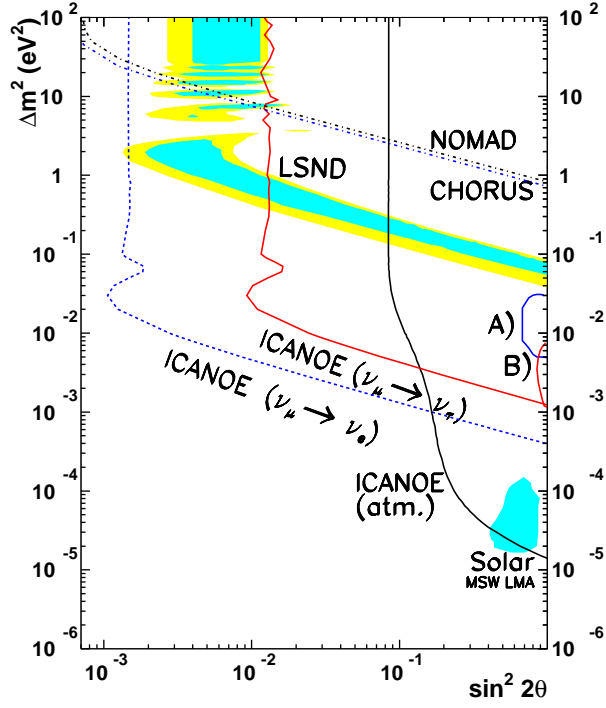,width=8cm}}
    \caption{Overview of the status of the
neutrino oscillations searches, displayed assuming two neutrino
mixing schemes in the $(\sin^22\theta, \dm2)$ plane. 
The 90\%C.L. allowed regions obtained 
from the Kamiokande (resp. Superkamiokande)
FC and PC samples are shown as A) (resp. B)).
The 90\% (resp. 99\%C.L.) regions consistent with the
LSND excess are shown as dark (resp. light) shaded areas
in the upper region of the plane. The shaded area
in the region $\dm2 \approx 10^{-5}\ \rm eV^2$
represents the large angle MSW solution
of the solar neutrino deficit.
CHORUS and NOMAD 90\%C.L. limits on $\numu\ra\nutau$ oscillations
are visible in the upper $\dm2$ region. The \ICANOE{}
sensitivities at 90\%C.L. are indicated by three curves: the
limit by direct observation of the atmospheric neutrinos
(``\ICANOE{} atm''); the direct tau appearance search at
the CNGS (``\ICANOE{} $\numu\ra\nutau$''); the direct electron
appearance search at the CNGS (``\ICANOE{} $\numu\ra\nue$''). 
}
\label{fig:overviewosc} 
\end{figure}

The sensitivity in the
classic $(\sin^22\theta,\dm2)$
plot is evidenced in Figure~\ref{fig:overviewosc}, for a data taking time of 4 years, 
with $4.5\times 10^{19}$ pot at each year.  We remark:
\begin{enumerate}
\item The recent results on atmospheric neutrinos 
(``A'' and ``B'' of Figure~\ref{fig:overviewosc}) can be thoroughly explored by 
appearance and disappearance experiments.  
For the current central value, both CNGS and cosmic ray data will give 
independent and complementary measurements and they will provide a 
precise $(\sin^22\theta,\dm2)$ determination.
\item In the mass range of LSND, the sensitivity is sufficient 
in order to solve definitely the puzzle.  
\item At high masses of cosmological relevance for
$\dm2< \approx 10\rm\ eV^2$, the sensitivity
to $\numu\ra\nutau$ oscillations is better or 
equal to the one of CHORUS and NOMAD.
\item In the atmospheric neutrino events, 
one can reach a level of sensitivity sufficient to detect also the 
effect until now observed in solar neutrinos.
This purely terrestrial detection of the LMA solar neutrino solution is
performed 
using neutrino in the GeV range, much higher than the one of solar
neutrinos.  
\end{enumerate}

Since we can observe and unambiguously identify both $\nue$ and
$\nutau$ components, 
the full (3 x 3) mixing matrix can be explored.  
By itself, this is one of the main justifications for the choice 
of the detector's mass.  

In the cosmic ray channel, all specific modes (electron, muon, NC)
are 
equally well observed without detector biases and down 
to kinematical threshold. The CR-spectrum being rather poorly known, 
a confirmation of the SuperKamiokande result requires detecting both (1) 
the modulation in the muon channel and (2) the lack of effect of 
the electron channel. The consistency of the simultaneous  
observation of the $L/E$ phenomenon in as many modes as they are 
available is a powerful tool in separating genuine flavour oscillations 
from exotic scenarios.  

In some favourable conditions, the direct appearance of the 
oscillated tau neutrino may be directly identified in the upgoing 
events, since even a few events will be highly significant. 

While in atmospheric neutrinos, 
the knowledge of the sign of the muon is of little 
relevance, in the case of the CNGS is a powerful tool to verify the 
neutrino nature after oscillation path, excluding for instance oscillation 
channels into anti-neutrinos.  

For a discussion on the nucleon decay searches, see Ref.~\cite{NNNnucdec}.


\section{Physics at the CNGS}

The design and performance of the CERN neutrino beam to Gran Sasso -
the CNGS facility - are described in a conceptual technical design
report \cite{NGSreport}. 
The CNGS beam performance for the new reference beam
are summarized in Table~\ref{tab:results1}. 
The rms radius of the $\nu_\mu$ CC event distribution is about 1.37\,km at 
Gran~Sasso.  
The 
expected numbers of detectable $\nu_\tau$ for 
$\sin^2 2\theta$\,=\,1 and a few typical values of
$\Delta m^2$ are shown in Table~\ref{tab:results2}. 

\begin{table}[tb]
\begin{tabular}{|c|c|c|} 
\hline
Energy region $E_{\nu_\mu}$ [GeV] & 1\,-\,30 & 1\,-\,100 \\ 
\hline
$\nu_{\mu}$\,[m$^{-2}$/pot]          &   7.1\,$\times$\,10$^{-9}$
&  7.45\,$\times$\,10$^{-9}$  \\
$\nu_{\mu}$ CC events/pot/kt     &
4.70\,$\times$\,10$^{-17}$ &
5.44\,$\times$\,10$^{-17}$ \\
\hline
\newcommand{\mean}[1] {\mbox{$ \left\langle #1 \right\rangle   $}}
$\mean{E}_{\nu_{\mu}\,fluence}$ [GeV] & & 17 \\
\hline 
fraction of other events:&  \multicolumn{2}{|c|}{}          \\
$\nu_e$/$\nu_{\mu}$                       &\multicolumn{2}{|c|}{0.8\,\%}
\\
$\overline{\nu}_{\mu}$/$\nu_{\mu}$        &\multicolumn{2}{|c|}{2.0\,\%}
\\
$\overline{\nu}_e$/$\nu_{\mu}$            &\multicolumn{2}{|c|}{0.05\,\%}
\\
\hline
\end{tabular}
\caption{Predicted performance of the new CNGS reference beam
for an isoscalar target.
The statistical accuracy of the Monte-Carlo simulations
is 1\,\% for the $\nu_{\mu}$ component of the beam,
somewhat larger for the other neutrino species.}
\label{tab:results1}
\end{table}

\begin{table}[htbp]
\begin{tabular}{|c|c|c|} 
\hline
Energy region $E_{\nu_\tau}$ [GeV] & 1\,-\,30 & 1\,-\,100 \\ 
\hline
 $\Delta m^2$\,=\,1\,$\times$\,10$^{-3}$\,eV$^2$  & 2.34  &  2.48 \\
 $\Delta m^2$\,=\,3\,$\times$\,10$^{-3}$\,eV$^2$ & 20.7  & 21.4 \\
 $\Delta m^2$\,=\,5\,$\times$\,10$^{-3}$\,eV$^2$ & 55.9  & 57.7 \\
 $\Delta m^2$\,=\,1\,$\times$\,10$^{-2}$\,eV$^2$ & 195  & 202 \\
\hline 
\end{tabular}
\caption{Expected number of $\nu_\tau$ CC events at Gran Sasso
per kton per year for an isoscalar target. 
Results of simulations for different values of $\Delta m^2$
and for $sin^2(2\theta)$\,=\,1 are given for 4.5\,$\times$\,10$^{19}$
pot/year. These event numbers do not take
detector efficiencies into account.}
\label{tab:results2}
\end{table}

Events will occur in the whole \ICANOE{} detector. 
As reference, we assume an exposure of $20\ \rm kton\times year$ 
for the liquid argon. This 
corresponds to four years running of the CNGS beam in shared mode.
For the events occuring in the solid detector, given the smaller
mass, the reference exposure is $10\ \rm kton\times year$.
The last three meters of the liquid target are defined as
a transition region, since beam events occuring in this region
are most likely to deposit energy in both targets.
Table~\ref{tab:nurates} shows the computed total event
rates for each neutrino species present in the beam for
the liquid, solid and in the transition region. 
Table \ref{tab:nurates} also shows, 
for three different values of $\Delta m^2$, the $\nu_\tau$ CC rates 
expected in case oscillations take place. 

\begin{table}[tb]
\begin{tabular}{||c|c|c|c|}
\hline
{\bf Process}  & liquid target  & transition & solid \\ 
\hline\hline 
 $\nu_\mu$ CC       &  54300 & 10200 & 27150 \\ 
 $\bar{\nu}_\mu$ CC     &       1090 & 200 & 545 \\ 
 $\nu_e$ CC     &  437  & 80 & 219 \\ 
 $\bar{\nu}_e$ CC   &          29 & 5 & 15 \\
 $\nu$ NC      &      17750 & 3330 & 8875 \\
 $\bar{\nu}$ NC      &       410 & 77 & 205 \\ \hline
\hline
 $\nutau$ CC, $\Delta m^2$ (eV$^2$)&  & &  \\ 
\hline\hline
 $1\times 10^{-3}$       &      52 & 10 & 26 \\
 $2\times 10^{-3}$       &      208 & 40 & 104 \\
 $3.5\times 10^{-3}$  &          620 & 115 & 310 \\ 
 $5\times 10^{-3}$  &          1250 & 235 & 625 \\ 
 $7.5\times 10^{-3}$  &         2850 & 535 & 1425 \\ 
 $1 \times 10^{-2}$  &         4330 & 810 & 2165 \\ \hline 
\end{tabular}
\caption{Expected event rates for an exposure of 20 kton$\times$ year
for the liquid target and 10 kton$\times$ year for the solid target. 
All the rates include nuclear corrections and are computed
for the proper target composition.
For standard processes, no oscillations is assumed.
For $\nu_\tau$ CC, we take two neutrino $\nu_\mu \to \nu_\tau$
with $\sin^22\theta=1$.}
\label{tab:nurates}
\end{table}

\subsection{Event kinematics and tau identification}
\label{sec:evkinetau}
Kinematical identification of the $\tau$ decay, which
follows the $\nutau$~CC interaction requires excellent
detector performance: good calorimetric features together
with tracking and event topology reconstruction
capabilities. The background from standard processes
are, depending on the decay mode of the tau lepton
considered, the $\nue$~CC events and/or the
$\numu$~CC and $\nu$~NC events.

In order to separate separate $\nutau$ events from the background,
two basic criteria, already adopted by the short baseline NOMAD
experiment, can be used:
\begin{itemize}
\item an unbalanced total transverse momentum due to neutrinos
produced in the $\tau$ decay,
\item a kinematical isolation of hadronic prongs and missing
momentum in the transverse plane.
\end{itemize}
In addition, given the baseline $L$ between CERN and GranSasso,
for the lower $\dm2$ values of the allowed region indicated by
the atmospheric neutrino results, we expect most of the oscillation
to occur at low energy. In this case, a criteria on the visible
energy is also very important to suppress backgrounds.

In order to apply the most efficient kinematic selection,
it is mandatory to reconstruct with the best possible
resolution the energy and the angle of the hadronic jet, with
a particular attention to the tails of the distributions.
Therefore, the energy flow algorithm should be designed with
care taking into account the needs of the tau search analyses.

A specially developed energy flow algorithm
has been tested on a sample of fully simulated
$\nue\ \rm CC$ events, in order to estimate the resolution
of the kinematical reconstruction on realistic events. 
It yields an average missing $P_T$ 
of $450\ \rm MeV/c$. This value improves to an average
of $410\ \rm MeV/c$ when the primary vertex is required
to lie within a fiducial volume of transverse dimensions
$7.8\times 7.8 m^2$. 

We used the neutrino data collected in the NOMAD detector
to probe the 
reliability of the physics simulation. 
$\nu_\mu$ CC events have been fully simulated and
reconstructed using NOMAD official packages. 
We found that the kinematics in the transverse plane are 
well reproduced by the
Monte-Carlo model. This is clearly not the case when nuclear corrections
are neglected.

\begin{table}[tb]
\begin{tabular}{||l|c|c|c|c|c|c|} \hline
Cuts & $\nu_\tau$ Eff. & $\nu_e$ & ${\bar{\nu}}_e$ & $\nu_\tau$ CC & 
$\nu_\tau$ CC & $\nu_\tau$ CC \\
 & ($\%$) & CC & CC & $\Delta m^2=$  &  
$\Delta m^2=$ & $\Delta m^2=$  \\
 &  &  &  & $10^{-3}$ eV$^2$ &  
$3.5 \times 10^{-3}$ eV$^2$ 
& $10^{-2}$ eV$^2$  \\ \hline
Initial                & 100 & 437 & 29  & 9.3 & 111 & 779    \\ 
Fiducial volume        &  88 & 383 & 25  & 8.2 & 97  & 686    \\ 
One candidate with     &     &     &     &     &     &        \\
momentum $> 1$ GeV     & 72  & 365 & 25  & 6.7 & 80  & 561    \\ 
$E_{vis} < 18$ GeV     & 67  &  64 & 5   & 6.2 & 75  & 522    \\ 
$P_T^e< 0.9$ GeV       & 54  &  31 & 3   & 5.0 & 60  & 421    \\ 
$P_T^{lep} > 0.3$ GeV  & 51  &  29 & 2   & 4.7 & 56  & 397    \\ 
$P_T^{miss}>0.6$ GeV   & 33  &   4 & 0.4 & 3.1 & 37  & 257    \\ \hline
\end{tabular}
\caption{Rejection of the
$\nue$~CC background in the
$\tau \to e$ analysis. Figures are normalized to an exposure
of 20 kton $\times$ year.}
\label{tab:taueana}
\end{table}

\subsection{$\numu\ra\nutau$ appearance searches}
\label{sec:appeartau}

The channel of tau decaying into an electron plus two neutrinos
provides the best sample for $\nutau$ appearance studies
due to the low background level. The intrinsic $\nue$,
$\bar\nue$ contaminations
of the beam amount to $\approx 470$ events for an exposure
of $20\ \rm kton\times year$. 

The comparison of this figure with the expected number of $\nutau$~CC events
decaying into electrons shows that the search of $\tau\ra e$ 
at the CNGS will have to be optimized {\it a posteriori}. Indeed
the $\nutau$ rate has a strong dependence on the exact value of the $\dm2$
in the parameter region suggested by the Super-Kamiokande data, and 
the $\dm2$ value is not well constrained by the atmospheric neutrino 
experiments.

For ``large'' values of $\dm2$, i.e. 
$\dm2>5\times 10^{-3}$, the rate of tau is spectacular and
exceeds the number of intrinsic beam $\nue$, $\bar\nue$~CC events,
i.e. $S/B > 1$ even prior to any kinematical cuts. So the
kinematical cuts can be very mild. An excess will be striking.

For our ``best'' value taken from atmospheric
neutrino results, i.e. $\dm2=3.5\times 10^{-3}\ \rm eV^2$,
the number of $\nutau~CC$ with $\tau\ra e$ is about 110, or
about a signal over background ratio of $110/470 \simeq 1/4$.
Here with modest kinematical cuts, we can extract 
statistically significant signals, as shown in the following
sections.

The most difficult region lies below $\dm2\approx 1.5\times
10^{-3}\ \rm eV^2$, for which, kinematical cuts
are tuned to suppress backgrounds by a factor more than 200
while keeping about half of the signal events.

\begin{table}[tb]
\begin{tabular}{||l|c|c|c|c|} \hline
Cuts & \multicolumn{4}{|c|}{$\nu_\mu$ NC} \\ \hline 
Initial                              & \multicolumn{4}{|c|}{17750}\\ 
Fiducial volume                      & \multicolumn{4}{|c|}{15550}   \\ 
\hline
 & Dalitz & $\gamma$ conv. & $\pi \to e$ & $\pi^\pm/\pi^0$  \\ 
One candidate                           &  275  & 4262 & 6.5  & 25  \\ 
$P_e> 1$ GeV                            &  79   & 1361 & 6.3  & 16  \\ 
$E_{vis} < 18$ GeV                      &  49   &  835 & 3.2  & 11  \\ 
$P_T^e< 0.9$ GeV                        &  46   &  794 & 1.8  &  9  \\ 
$P_T^{lep} > 0.3$ GeV                   &  24   &  429 & 1.7  &  8  \\ 
$P_T^{miss}>0.6$ GeV                    &  19   &  350 & 1.3  &  7  \\ 
Imaging and $dE/dx$                     & $<1$  & $<1$ & $<1$ & $<1$ \\ \hline
\end{tabular}
\caption{ $\nu_\mu$ NC background to the 
$\tau \to e$ analysis. Results are normalized to an exposure 
of 20 kton $\times$ year. We illustrate background reduction
by means of kinematical criteria only. Imaging and $dE/dx$ measurements reduce the
NC background to a negligible level.}
\label{tab:numuncback}
\end{table}

In the following paragraphs, we discuss background sources
and their suppression.

{\bf $\nue~CC$ rejection:}
The main background from genuine leading electrons comes from 
the CC interactions
of the $\nu_e$ and $\bar{\nu}_e$ components of the
beam. 
In Table \ref{tab:taueana} we summarize the list of sequential cuts 
applied to reduce the $\nu_e$ and $\bar{\nu}_e$ CC backgrounds
and the expected number of signal events for three different $\Delta
m^2$ values. 
The most sensitive analysis predicts, for a 20 kton $\times$ year exposure, 
a total background of 4.4 events for a total $\tau$ selection
efficiency of 33$\%$. 

\begin{figure} 
\mbox{\epsfig{file=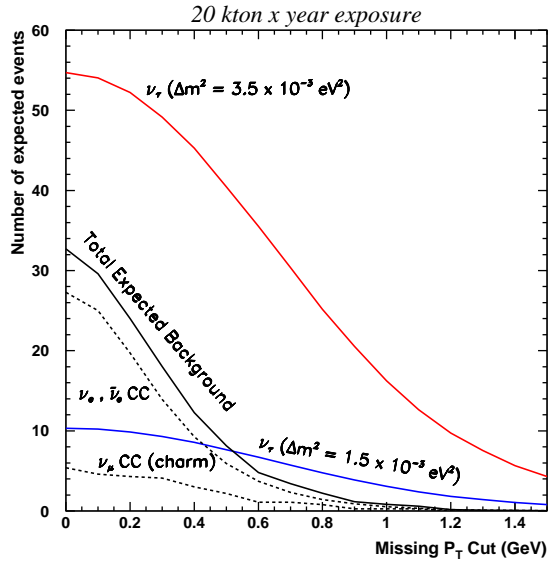,width=8.cm}} 
\caption{Number of $\nu_e, \ \bar{\nu}_e$ background and tau events
expected for a 20 kton $\times$ year exposure as a function of the
missing transverse momentum.}
\label{fig:ptmisscut}
\end{figure}

{\bf $\nu~NC$ rejection:}
Neutral current events contribute to the background
from four sources: (1) electrons from Dalitz decays,
(2) early photon conversions, (3) interacting charged pions
and (4) $\pi^\pm/\pi^0$ overlap.
Table~\ref{tab:numuncback} summarizes the rejection power of
kinematics criteria for the four sources that contribute to $\nu~NC$
background. The requirement on the electron
candidate energy $E_e>1\ \rm GeV$ suppresses about one third of the
Dalitz, pion overlap and $\pi^0$ conversions induced backgrounds, 
since electrons in the jet are soft.

The ultimate discrimination of these backgrounds relies primarily on the
imaging capabilities and on $dE/dx$ measurements. 
The combination of $dE/dx$ information together with 
kinematics criteria is sufficient to reduce $\nu~NC$ background 
to a negligible level.

{\bf $\numu~CC$ rejection:}
Charged current events can contribute to the background
in a similar way as the neutral current events described
above when the leading muon escapes detection. 
In case the muon is not identified, the event will appear
in first instance as a neutral current event. The source
of electrons which can induce backgrounds are then similar
to those discussed previously and are reduced to a negligible level for reasons 
already discussed.
A more important source of background specific to charged
current interactions comes from the 
decays of charmed mesons. At the CNGS energies 
$\sigma (\nu_\mu N \to \mu \ c \ X)/\sigma (\nu_\mu N \to \mu X)
\approx 4\%$, therefore for a total exposure of 20 kton $\times$ year
we expect to collect about 200 events where a charmed meson decays into 
a positron and a neutrino. These events resemble kinematically
the real $\nu_\tau$ events,
since they have a neutrino in the final state and possess 
a softer energy spectrum and a genuine sizeable missing transverse momentum. 
After all cuts, the expected number of charm induced background events
$n^b_{CC \ (charm)}$ for a total exposure of 20 kton $\times$ year is
at the level of 1 event.

\begin{table}
\begin{tabular}{||c|c||c|c|c|} \hline
$\Delta m^2$ (eV$^2$)& $\nu_\tau$ CC & $\nu_e , \ {\bar{\nu}}_e$ CC & $\nu_\mu, \
{\bar\nu}_\mu$ CC & $\nu_\mu$ NC \\ \hline 
$1 \times 10^{-3}$                & 3 & & & \\\cline{1-2}
$2 \times 10^{-3}$                & 12 & & & \\\cline{1-2}
$3 \times 10^{-3}$                & 26 & & & \\\cline{1-2}
{\boldmath $3.5 \times 10^{-3}$}  & {\bf 35} & {\bf 4.1}& {\bf 1.0} & {\boldmath $<1$}\\\cline{1-2}
$5 \times 10^{-3}$                & 71 & & & \\\cline{1-2}
$7 \times 10^{-3}$                & 121 & & & \\\cline{1-2}
$1 \times 10^{-2}$                & 248 & & & \\ \hline
\end{tabular}
\caption{$\tau \to e$ analysis summary. For a total exposure of 20 $\times$ year 
we show the expected number of $\tau$ events for different $\Delta m^2$ values. 
The last three columns show the expected background.}
\label{tab:finaltaue}
\end{table}

\subsection{Combined $\numu\ra\nutau$ sensitivity}

Table~\ref{tab:finaltaue} summarizes the expectations for the $\tau
\to e$ analysis once kinematics criteria and muon vetoes have
been applied to every potential background source. In conclusion,
we obtain for a 20 kton $\times$ year exposure, that the overall
electron selection efficiency is 32$\%$ for an expected number of
about five background events. The expected number of fully
identified tau events at the central 
$\Delta m^2$ value of $3.5 \times 10^{-3}$ eV$^2$ is 35.
 
We show in Figure~\ref{fig:ptmisscut} 
as a function of the cut on the missing transverse momentum, the 
number of expected background and tau events for two different 
$\Delta m^2$ values. We see that even for a value as low as 
$1.5 \times 10^{-3}$ eV$^2$, a $P_T^{miss}$ cut above 0.6 GeV 
gives a $S/B$ ratio in excess of 1.

\begin{figure} 
\mbox{\epsfig{file=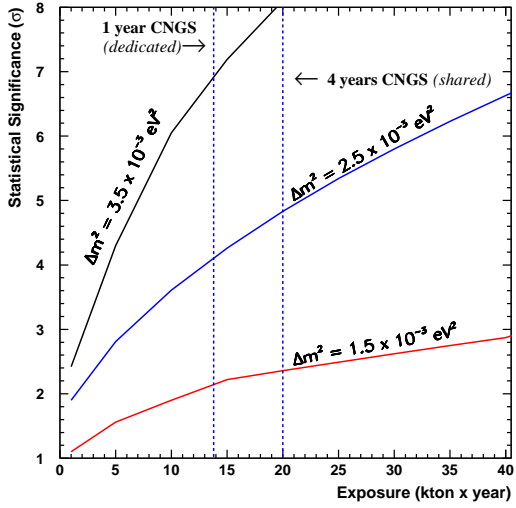,width=7.5cm}} 
\caption{$\tau$ appearance statistical significance for different 
$\Delta m^2$ values as a function of the exposure. The arrows indicate 
the statistical significance achieved for one year of dedicated run of 
the CNGS and four years of CNGS running in shared mode.}
\label{fig:sigma}
\end{figure}

Finally it is crucial to study the exposures needed to obtain a statistically
significant $\tau$ appearance for different neutrino 
oscillation scenarios. We see in Figure \ref{fig:sigma} 
that for $\Delta m^2= 3.5$ eV$^2$, few months of data 
taking will suffice to claim a 3$\sigma$ effect. However for 
$\Delta m^2$ values of about $10^{-3}$ eV$^2$, exposures above 20 kton 
$\times$ year are needed to obtain an effect in excess of 2$\sigma$.
We conclude that after four years of running of CNGS in shared mode 
or after one year of running in dedicated mode, the \ICANOE{} detector 
will observe a statistically significant 
$\nu_\mu \to \nu_\tau$ oscillation signal for most of the $\Delta m^2$ 
values presently favored by atmospheric neutrino data.

\subsection{$\numu\ra\nue$ oscillation search}
\label{sec:appeare}

The unambiguous detection and identification of $\nu_e$ CC events
endows \ICANOE{} with the ability of performing also a 
$\nu_\mu \to \nu_e$ oscillation search. In this case, the oscillation
reveals itself as an excess on the number of expected events having
a leading identified electron. Given the expected rates for a 20 kton
$\times$ year exposure, the statistical error is about
4$\%$. Therefore the sensitivity to $\nu_\mu \to \nu_e$ oscillations
is dominated by the systematic error on the beam knowledge.

\begin{figure} 
\mbox{\epsfig{file=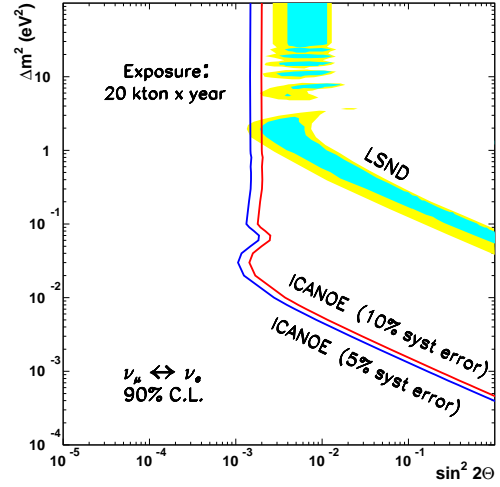,width=7.cm}} 
\caption{\ICANOE{} 90$\%$ C.L. exclusion region in case no $\nu_\mu \to
\nu_e$ are experimentally observed.}
\label{fig:mucont}
\end{figure}

Figure \ref{fig:mucont} shows the 90$\%$ C.L. contours in case no
oscillations are observed assuming overall systematic errors of 
5$\%$ and 10$\%$. We observe that nearly the whole region favored by
the LSND claim is comfortably covered. The excluded values are 
$\sin^2 2\theta > 2 \times 10^{-3}$ at high $\Delta m^2$ and 
$\Delta m^2 > 4 \times 10^{-4}$ eV$^2$ for maximal mixing.

\section{Atmospheric neutrinos}
The physics goals of the new atmospheric neutrino measurements 
are to firmly establish
the evidence of neutrino oscillations with a different
experimental technique, possibly free of systematic
biases, measure the oscillation parameters and
clarify the nature of the oscillation mechanism.
\ICANOE{} will provide, in addition
to comfortable statistics, an observation of atmospheric neutrinos
of a very high quality.
Unlike measurements obtained up to 
now in Water Cherenkov detectors, which are in practice limited
to the analysis of ``single-ring'' events, complicated final
states with multi-pion products, occurring mostly at energies
higher than a few GeV, will be completely analyzed and reconstructed
in \ICANOE. This will be a significant improvement with respect to previous 
observations.

We have considered the following three methods:
\begin{itemize}
\item {\bf $\numu$ disappearance}: detection of the oscillation pattern in
the $L/E$ distribution, where $L$ is the neutrino
pathlength and $E$ its energy;
\item {\bf $\nu_\tau$ appearance} : comparison of the NC/CC with expectation;
\item {\bf direct $\nu_\tau$ appearance} : 
comparison of upward and downward rates of ``tau-like'' events.
\end{itemize}
together with the well established ones:
\begin{itemize}
\item {\bf the double ratio, $(\numu/\nue)_{obs}/(\numu/\nue)_{MC})$};
\item {\bf up/down asymmetry};
\end{itemize}

The tau appearance measurements can shed light on the
nature of the oscillation mechanism, by discriminating between the
hypothesis of oscillations into a sterile or a tau neutrino. 
The $\nu_\tau$ appearance method is based on $\nu_{\tau}$ CC
interactions with $\nu_\tau$ decaying into hadrons, hence 
to ``neutral-current-like'' events of high
energy. An excess of ``NC-like'' events from the bottom will indicate
the presence of oscillation to the $\nu_\tau$ flavour.
A kinematical analysis of the final state particles in the event 
can be used to further improve the statistical significance of the excess. 
Such a feature can only be obtained in a detector with the resolution
of the \ICANOE{} liquid target, in which all final state particles can be 
identified
and precisely measured. The kinematical method would
allow the evidence for ``tau-like'' events in the atmospheric neutrino
beam. 

Both the $\nu_\mu$ disappearance and the direct $\nu_\tau$ appearance
methods are weakly depending on the predictions of neutrino
event rates, since they rely on the comparison of rates induced by a
downward going and upward going neutrinos. 

The $NC/CC$ method, already investigated by SuperKamiokande, can be
significantly improved compared to this latter measurement.
In \ICANOE,
imaging in the liquid target provides a clean bias free identification
of neutral-current, independent on the hadronic final state,
since the identification is based on the absence of an electron
or a muon in the final state.

In the following sections, we will study our 
results for three different exposures:
5 kton$\times$year corresponding to 1 year of operation,
20 kton$\times$year for 4 years and an ultimate exposure
of 50 kton$\times$year or 10 years of operation.

The computed rates for the different neutrino processes 
(in events/kton/year) and their mean energies
are quoted in Table~\ref{tab:FLUX_RATES}, using
the FLUKA-3D atmospheric neutrino fluxes\cite{1d3d}.

\begin{table}[tb]
\begin{tabular}{|c|c|c|c|c|c|}
\hline
& & & & &  \\
Process & elastic & single-$\pi$ & inelastic & Total & $<$E$>$ \\
& & & & & (GeV)   \\
\hline
$\nu_{\mu}\rm\ CC$       & 66.7 & 15.9 & 24.4 & 107.0 & 2.36 \\
$\bar{\nu}_{\mu}\rm\ CC$ & 12.2 &  5.3 &  9.8 &  27.2 & 3.34 \\
$\nu_e\rm\ CC$          & 39.4 &  8.4 & 12.1 &  59.9 & 1.60 \\
$\bar{\nu}_e\rm\ CC$    &  5.4 &  2.1 &  4.2 &  11.7 & 2.36 \\
$\nu\rm\ NC$            & 42.9 &  8.6 & 13.2 &  64.8 & 1.94 \\
$\bar{\nu}\rm\ NC$      & 21.1 &  3.5 &  5.0 &  29.6 & 2.00 \\
\hline
\end{tabular}
\caption{Expected atmospheric neutrino rates per kton$\times$year
on Argon in case of no oscillations.  The figures have been computed
with \FLUKA{} 3D atmospheric neutrino flux with geomagnetic cutoffs and
include all nuclear effects. All NC events are included, even though only a
fraction of quasi-elastic ones will end up in an observable final state.}
\label{tab:FLUX_RATES}
\end{table}

\begin{figure}[tb]
\mbox{
\epsfig{file=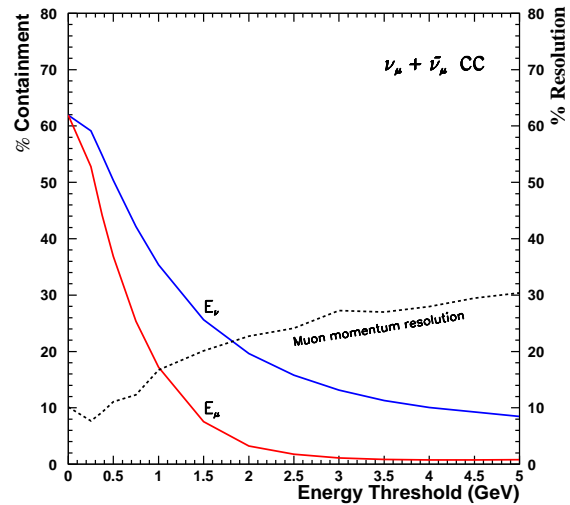,width=7.5cm}}
\caption{Integral distributions showing the containment for
$\nu_{\mu}$ CC as a function of the neutrino energy and
the leading muon momentum (solid lines). Differential distribution
showing the muon momentum resolution as a function of muon momentum
(dashed line), including both contained and partially contained events.}
\label{fig:CONT-MUO}
\end{figure}

\subsection{Event containment and muon measurement}
The muon measurement is crucial to most atmospheric neutrino analyses. In 
\ICANOE{}, we achieve the required performances using the multiple 
scattering measurement rather than resorting to a high--density, coarser
resolution detector.
Keeping a low density detector, high granularity detector 
imaging allows in addition the
identification and measurement of electrons and individual hadrons in
the event.

``Fully contained events'' are those for which the visible products
of the neutrino interaction are completely contained within
the detector volume. 
``Partially contained events'' are $\numu\rm\ CC$ events
for which the muon exits the detector volume (only muons
are penetrating enough).

Figure~\ref{fig:CONT-MUO} shows containment of charged
current events for different incoming
neutrino energies or muon momentum thresholds. 
Clearly, because of the average energy loss of the muon in 
argon (about $210\rm \ MeV/m$ for a m.i.p.), muons produced 
in neutrino events are often energetic
enough to escape from the subdetector volume.

It should first be noted that for contained events the muon energy resolution
is $4\%$ from $dE/dx$ measurements. For escaping muons, the high
granularity of the imaging allows to collect a very precise determination
of the track trajectory. Therefore the multiple scattering method
can be effectively used to estimate the momentum
of the escaping muons. This method requires in practice tracks in excess of
1~meter and works extremely well in the relevant energy range of
atmospheric neutrino events (typically below 10~GeV). 

The average muon momentum resolution as a function
of the energy threshold is shown in Figure~\ref{fig:CONT-MUO}.
This resolution has been computed using the range measurement for contained
muons and multiple scattering method for the escaping ones.
For energies below 1~GeV, the average muon momentum resolution is about 10\%.
It increases slowly as a function of the muon momentum and reaches
about 30\% at 5~GeV.


\subsection{Incoming neutrino angular resolution}

\begin{figure}
\mbox{
\epsfig{file=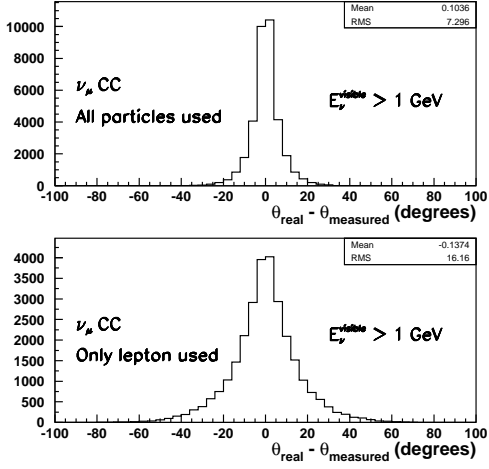,width=7cm}
}
\caption{Zenith angle resolution. The top plot shows the resolution
obtained by reconstructing the incoming neutrino direction using all
particles momenta, the bottom plot shows the resolution obtained using
only leading lepton momentum.}
\label{ZEN-RES}
\end{figure}

The reconstruction of the zenith angle of the incoming $\nu$ is of
great importance in the search of oscillations in atmospheric
neutrinos. \ICANOE{} allows
for a good
reconstruction of the incoming neutrino variables (i.e. incidence
angle, energy) by using the information coming from all particles
produced in the final state.
Figure~\ref{ZEN-RES} shows the distribution of the difference 
between the real and reconstructed neutrino angle 
for events with $E_{\nu}>1$~GeV.  
The improvement on the angular resolution is visible.
The RMS of the distribution improves from $\sim 16$ to $\sim 8$ degrees
after the inclusion of the hadronic jet in the reconstruction.

\subsection{The flavor ratio and up/down asymmetry}

Given the clean event reconstruction of \ICANOE, 
the ratio $R$ of
``muon-like'' to ``electron-like'' events can be determined
free of experimental systematic errors. In fact, the expected
purity of the samples is above 99\%. In particular,
the contamination from $\pi^0$ in the
``electron-like'' sample is expected to be completely negligible.
The measurement accuracy will be dominated
by the statistical uncertainty and by the theoretical systematic error on
the double ratio.

In order to estimate statistical sensitivities, 
we show in Table~\ref{MUEALLRATES} 
the values and statistical errors of $R$
for different exposures, assuming an oscillation $\nu_{\mu}
\to \nu_{x \neq \mu}$, with parameters $\sin^2 2\theta =  1$ and 
$\Delta m^2 = 3.5 \times 10^{-3}\rm\ eV^2$. 
The table lists the expected results when
using all events or only fully contained events
which turn out to be quite similar. 

Clearly, after an exposure corresponding to about four years of running
of \ICANOE, the statistical error reaches a level below 5\%. We expect that
further  theoretical improvements should reduce the systematic error 
to a level matched to statistical precision achievable
in \ICANOE.

\begin{table}[h]
\begin{tabular}{|c|c|c|}
\hline
Exposure (kton$\times$year) & \multicolumn{2}{c|}{$R$} \\
& all events & contained \\
\hline
 5 & $0.696 \pm 0.067$ & $0.674\pm 0.074$ \\
\hline
20 & $0.696 \pm 0.033$ & $0.674\pm 0.037$ \\
\hline
50 & $0.696 \pm 0.021$ & $0.674\pm 0.023$ \\
\hline
\end{tabular}
\caption{$R$ as a function of the exposure assuming maximal mixing 
and $\Delta m^2 = 3.5 \times 10^{-3}$ eV$^2$. In left column: results
obtained by using all events; right column: results obtained by using
only fully contained events. Quoted errors are of statistical
nature. }
\label{MUEALLRATES}
\end{table}

\begin{figure}[tb]
\centering
\mbox{
\epsfig{file=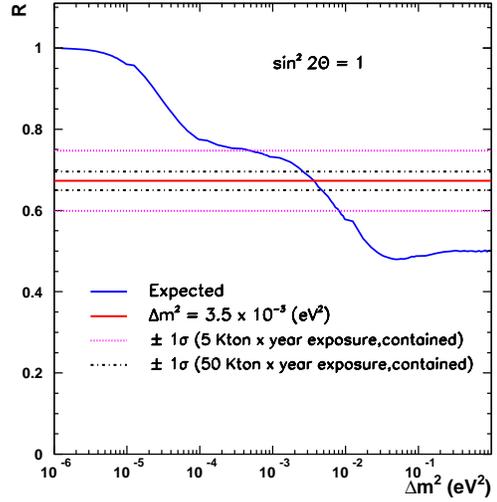,width=7.5cm}
}
\caption{Expected $R$ value as a function of $\Delta m^2$. Assuming a
true $\Delta m^2$ value of $3.5 \times 10^{-3}$ eV$^2$, the $68\%$
confidence intervals are given for 5 and 50 kton$\times$year.}
\label{MUECONTDET}
\end{figure}

\begin{figure*}[tb]
\mbox{
\epsfig{file=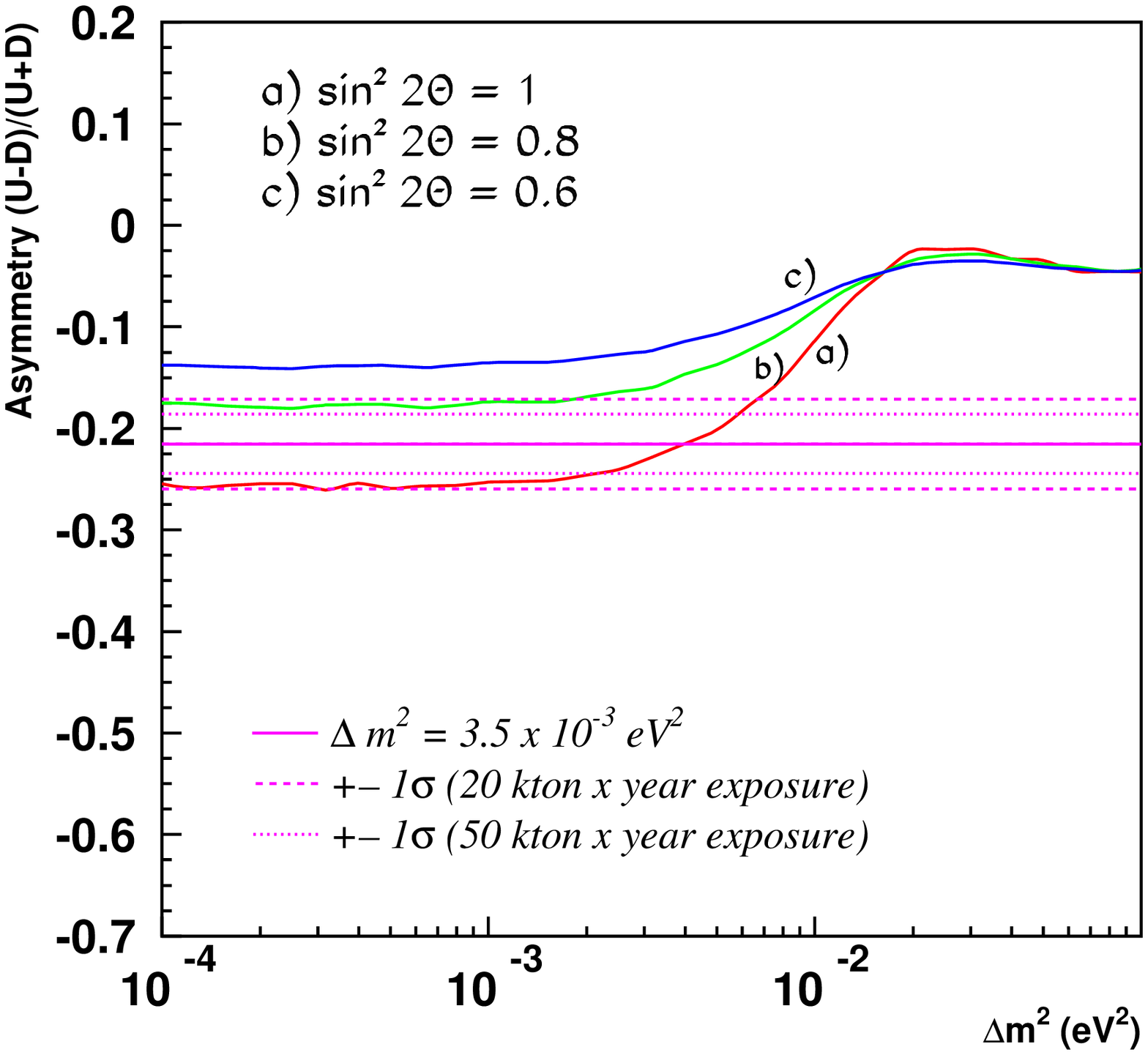,width=8cm}
\epsfig{file=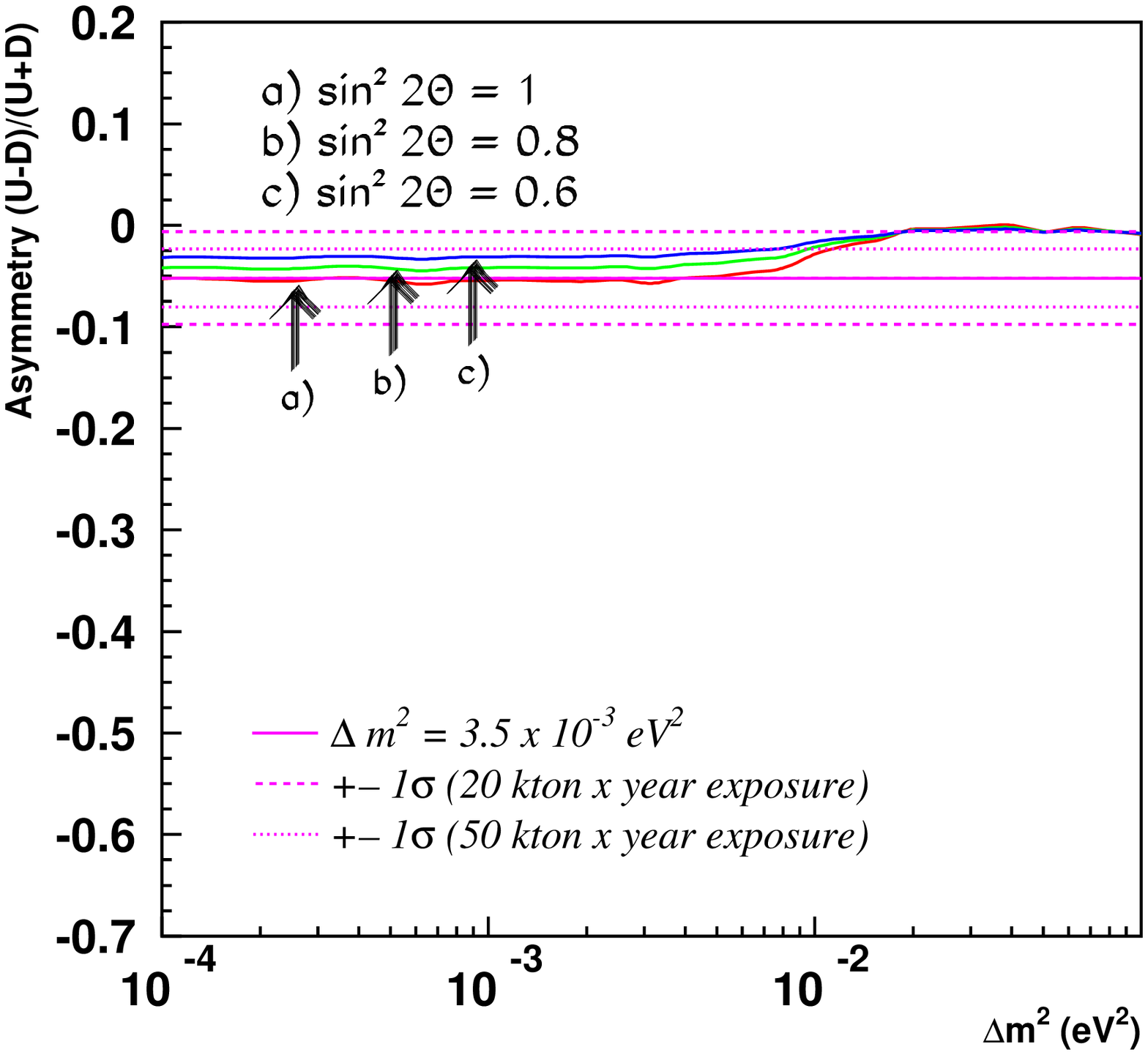,width=8cm}
}
\caption{
Expected Up-Down asymmetry for $\sin^2(2\theta)=1,0.8,0.6$ as a function 
of $\dm2$ for low momentum
(P$\le$0.4~GeV) $\nu_\mu$ events. (left) when all
particles are used to reconstruct the incoming neutrino
direction (right) only the lepton is used. The error bands show the $1~\sigma$
uncertainty for 20~kton$\times$year and 50~kton$\times$year exposures
and a 5\% systematic error assuming $\sin^2 2\theta =  1$ and 
$\Delta m^2 = 3.5 \times 10^{-3}\rm\ eV^2$.} 
\label{fig:UPDNASSY1}
\end{figure*}

From the value of $R$ and from its zenith angle dependence we can
obtain the allowed parameter regions of neutrino oscillations. Figure
\ref{MUECONTDET} shows how precisely we can determine $\Delta m^2$ in
the oscillation case. The $1\sigma$ regions corresponding to 5 and 50
kton$\times$year have been computed using contained events only.

The expected up/down asymmetries are shown in
Figure~\ref{fig:UPDNASSY1} for 
three different mixing angles as a function of $\Delta m^2$.
Note in Figure~\ref{fig:UPDNASSY1} 
the much better asymmetry resolution of \ICANOE{} for low energy muons
when compared to a measurement including the lepton only.


\subsection{$\numu$ disappearance -- $L/E$ studies}
\label{sec:lovere}
In order to verify that atmospheric neutrino disappearance
is really due to neutrino oscillations, an effective method 
consists in observing the modulation 
given by the characteristic oscillation probability:
\begin{equation}
P\left(\frac{L}{E}\right)=1-\sin^2(2\theta)\sin^2\left(1.27\Delta m^2 \frac{L}{E}\right)
\end{equation}
with $L$ in km, $E$ in GeV, $\Delta m^2$ in eV$^2$. 
This modulation will be characteristic of a given $\Delta m^2$,
when the event rate is plotted as a function of the
reconstructed $L/E$ of the events when compared to theoretical
predictions. The ratio of the observed and predicted spectra
has the advantage of being
quite insensitive to the precise knowledge of 
the atmospheric neutrino flux, since the oscillation 
pattern is found by dips in the $L/E$ distribution
while the neutrino interaction spectrum is known to be a 
slowly varying function of $L/E$.
Such a method is in principle capable of measuring $\Delta m^2$
exploiting atmospheric neutrino events.

A smearing of the modulation is introduced by the finite $L/E$ resolution
of the detection method.
Precise measurements of energy and direction of both the muon 
and hadrons are therefore
needed in order to reconstruct precisely the neutrino
$L/E$. This is quite well achieved in \ICANOE. The contained muons
can be measured with a resolution of 4\%, while the non-contained
muons are measured by multiple scattering method .

The RMS reconstructed $L/E$ resolution is about 30\%
for events with $E_{visible}>1\rm\ GeV$.

\begin{figure}[tb]
\mbox{
\epsfig{file=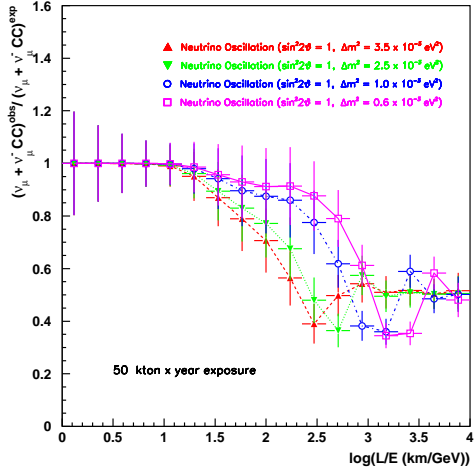,width=7cm}
}
\caption{Survival probability as a function of the $L/E$ ratio assuming 
neutrino oscillation hypothesis and for various $\Delta m^2$ values and
for 50 kton$\times$year. Only statistical error has been considered.}
\label{L_OVER_E_DM2}
\end{figure}

The $\nu_{\mu}$ survival probability as a function of $L/E$ let
us determine the value of $\Delta m^2$ in case of oscillation is
confirmed. In figure \ref{L_OVER_E_DM2} we can see the survival
probabilities of $\nu_{\mu}$ for neutrino oscillation hypothesis and
four different values of $\Delta m^2$. The first minimum on the
survival probability happens at highest $L/E$ values for the lowest
$\Delta m^2$ values, and allows us to discriminate between them for an 
exposure of 50 kton$\times$year.

The most favored solution of the atmospheric neutrino anomaly
is through $\numu\ra\nutau$ oscillations.
However, alternative 
explanations, like neutrino decay, cannot yet be excluded~\cite{decay}.
For example,  in a model
in which one of the mass-eigenstates of neutrinos with $\nu_{\mu}$ flavour 
content decays, the disappearance probability can be
described by the expression:
\begin{equation}
P(\nu_{\mu} \to \nu_{x \ne \mu}) = (\sin^2\theta + \cos^2\theta
e^{-\alpha L/2E})^2.
\end{equation}
Such a model gives an equally good fit
for the choice of parameters $\alpha = 1/63$ GeV/km, 
$\cos^2 \theta = 0.30$~\cite{decay}.

The capability of distinguishing between the two hypothesis
 depends on the resolution in measuring the $L/E$
ratio, which depends on the angular and momentum resolution. 

\begin{figure}[tb]
\mbox{
\epsfig{file=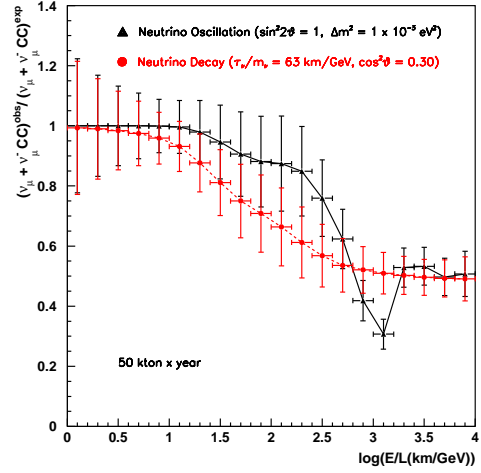,width=7cm}
}
\caption{Survival probability as a function of the $L/E$ ratio for
oscillation (triangles) and decay hypothesis (circles). Only
statistical error has been considered.}
\label{L_OVER_E_COMP}
\end{figure}

Figure \ref{L_OVER_E_COMP} shows the survival
probabilities as a function of $L/E$  for the neutrino 
decay hypothesis with $\alpha =
m_{\nu}/\tau_{\nu} = 1/63$ GeV/km and $\cos^2\theta = 0.30$, and the
oscillation hypothesis with $\sin^2 2 \theta = 1$ and $\Delta m^2 =
1.0 \times 10^{-3}$, for an exposure of 50 kton$\times$year. Both
hypothesis are distinguishable from each other at around 2000 km/GeV
within the statistical errors.

\subsection{(Direct) appearance of tau neutrinos}

To discriminate between $\nu_{\mu} \to \nu_{\tau}$ and $\nu_{\mu} \to
\nu_s$ oscillations, we measure the ratio $R_{NC/e} =
\frac{NC^{obs}/\nu_e CC^{obs}}{NC^{exp}/\nu_e CC^{exp}}$. An
oscillation to an active neutrino leads to $R_{NC/e} =1$, while
$R_{NC/e} \sim 0.7$ is expected for an oscillation to sterile
neutrino.

\begin{table}[h]
\begin{tabular}{|c|c|c|}
\hline
Exposure (kton$\times$year) & \multicolumn{2}{c|}{$R_{NC/e}$} \\
& all events & contained \\
\hline
 5 & $ 0.674 \pm 0.086$ & $0.670 \pm 0.087$ \\
\hline
20 & $ 0.674 \pm 0.043$ & $ 0.670 \pm 0.044$ \\
\hline
50 & $ 0.674 \pm 0.027$ & $0.670 \pm 0.028$ \\
\hline
\end{tabular}
\caption{$R_{NC/e}$ as a function of the exposure assuming oscillation 
to a sterile neutrino. Quoted errors are of statistical nature.}
\label{NCEALLSTERILE}
\end{table}

Table \ref{NCEALLSTERILE} shows values and errors of $R_{NC/e}$ in
case of oscillation to a sterile neutrino, for all events and fully
contained events respectively.


For $\Delta m^2 \leq 10^{-2}$ eV$^2$, oscillations of $\numu$ into
$\nutau$ would in fact result in an excess of ``neutral-current-like'' events
produced by upward neutrinos with respect to downward, since 
charged-current $\nutau$ interactions would contribute to the
``neutral-current-like'' event sample, due to the large $\tau$ branching 
ratio into
hadronic channels. Moreover, due to threshold effect on $\tau$
production, this excess would be important at high energy. 
Oscillations into a sterile neutrino would instead result in a     
depletion of upward muon-less events. Discrimination between
$\numu \to \nutau$ and $\numu \to \nu_{s}$ is thus obtained from a
study of the asymmetry of upward to downward muon-less events.  
Because this method works with the high energy component of
atmospheric neutrinos, it becomes effective for relatively large
values of $\Delta m^2$ ($\ge 3 \times 10^{-3}$ eV$^2$). 

Charged current $\nu_\tau$ rates for
five $\Delta m^2$ hypothesis: $5\times 10^{-4}$,
$1\times 10^{-3}$, $3.5\times 10^{-3}eV^2$, $5\times 10^{-3}eV^2$
and $1\times 10^{-2}eV^2$ are listed in Table~\ref{tab:taurate}. We
see that the rates saturate at about one event per kton$\times$year
for the larger $\Delta m^2$ values. Such small rates pose a major 
experimental challenge in the detection of $\nu_\tau$ in the
cosmic ray induced neutrino flux. 

\begin{table}
\begin{tabular}{||r|c|c|c||c|c|}\hline\hline
\multicolumn{4}{||c||}{$\nu_\tau + \bar{\nu}_\tau$ CC (NUX, Fluka 3D flux)} 
& Rel. to & Rel. to
\\\cline{1-4}
 & \multicolumn{3}{|c||}{Rate (kton$\times$ year)} & Fluka 1D &
Bartol \\\cline{2-4}
$\Delta m^2$ (eV$^2$)  & DIS & QE & Sum & & \\ \hline
$5 \times 10^{-4}$   & 0.11 & 0.11 & 0.22 & 0.96 & 0.81 \\
$1 \times 10^{-3}$   & 0.28 & 0.18 & 0.46 & 1.02 & 0.84 \\
$3.5 \times 10^{-3}$ & 0.59 & 0.21 & 0.80 & 1.00 & 0.81 \\
$5 \times 10^{-3}$   & 0.64 & 0.24 & 0.88 & 1.01 & 0.80 \\
$1 \times 10^{-2}$   & 0.70 & 0.20 & 0.90 & 0.99 & 0.78 \\ \hline
\end{tabular}
\caption{Expected $\nu_\tau$, 
$\bar{\nu}_\tau$ absolute rates for five different $\Delta m^2$
with FLUKA-3D fluxes and relative to FLUKA-1D and Bartol fluxes.}
\label{tab:taurate}
\end{table}

The total visible energy ($E_{visible}$) is a suitable 
discriminant variable to enhance the $S/B$ ratio. After cuts, 
surviving events are classified as: $n_b$ (number of expected 
downward going background) and 
$n_t = n_b + n_s$ (number of expected upward going events, 
where $n_s$ is the number of taus). 
The statistical significance of the expected $n_s$ excess is evaluated 
following two procedures: 
\begin{itemize}
\item The $f_b$ and $f_t$ pdf's are integrated over the whole 
spectrum of possible measured $r$ values and the overlap between 
the two is computed: 
$P_\alpha \equiv \int_0^\infty min(f_b(r), \ f_t(r)) dr$,
where $f_b$ and $f_t$ are the Poisson p.d.f.'s 
for means $\mu=n_b$ and $\mu=n_t$ respectively.
The smaller the overlap integrated probability ($P_\alpha$) 
the larger the significance of the expected excess.
\item computing the probability $P_\beta 
\equiv \int_{n_t}^\infty \frac{e^{-n_b} n_b^r}{r!} dr$ that, due to a
statistical fluctuation of the unoscillated data, 
we measure $n_t$ events or more when $n_b$ are expected. 
\end{itemize}

For a 50 kton$\times$year 
exposure, the results of a search based on $E_{visible}$ 
are shown in Table \ref{tab:hadron}.
We see that a cut on
visible energy between 6 and 7 GeV results in: 
(1) an overlap integrated probability between the two distributions 
amounting to $25-26\%$.
(2) a Poisson 
probability that the
measured excess (``$\tau$ bottom'') corresponds to a statistical 
fluctuation is $0.6 - 0.8 \%$.

\begin{table}[htb]
\begin{tabular}{||l|c|c||c|c|} \hline\hline
\multicolumn{5}{||c|}{ {\bf 50 kton$\times$year exposure}} \\\hline
$E_{Visible} cut$ & $\nu$ NC top & $\tau$ bottom & $P_\alpha$
($\%$) & $P_\beta$ ($\%$) \\ \hline
$ > 1 $ GeV & 327 & 22 & 55.0 & 10.8 \\
$ > 2 $ GeV & 150 & 22 & 38.6 & 3.54 \\
$ > 3 $ GeV &  95 & 21 & 30.6 & 1.6 \\
$ > 4 $ GeV &  67 & 20 & 25.3 & 0.8 \\ 
$ > 5 $ GeV &  51 & 17 & 27.3 & 0.9 \\
$ > 6 $ GeV &  40 & 16 & 24.6 & 0.6 \\
$ > 7 $ GeV &  33 & 14 & 26.6 & 0.8 \\  
$ > 8 $ GeV &  28 & 13 & 26.7 & 0.8 \\
$ > 9 $ GeV &  23 & 12 & 26.2 & 0.7 \\     
$> 10 $ GeV &  21 & 11 & 28.3 & 0.9 \\ \hline     
\end{tabular}
\caption{Number of NC and tau events as a
function of the visible energy cut. The statistical sample used
corresponds to an exposure of 50 kton$\times$year.}
\label{tab:hadron}
\end{table}

The search for $\nu_\tau$ appearance can be improved taking advantage
of the special characteristics of $\nu_\tau$ CC and the subsequent
decay of the produced $\tau$ lepton when compared to CC and NC interactions 
of $\nu_\mu$ and $\nu_e$, i.e. by making use of $\vec{P}_{lepton}$ 
and $\vec{P}_{hadron}$.

The information related to the directionality of the incoming
neutrino (i.e. the beam direction!) is missing. As a result, 
we have three kinematical independent variables in order to
separate signal from background. 
After a careful evaluation of the performance of different
combinations of variables, we decided to use:
$E_{visible}$,
$y_{bj}$ (the ratio between the total hadronic energy and
$E_{visible}$), and
$Q_T$ (the transverse momentum of the $\tau$
candidate with respect to the total measured momentum)
which contains the information on the isolation 
of the tau candidate from the recoiling jet.

\begin{figure}[tb]
\mbox{\epsfig{file=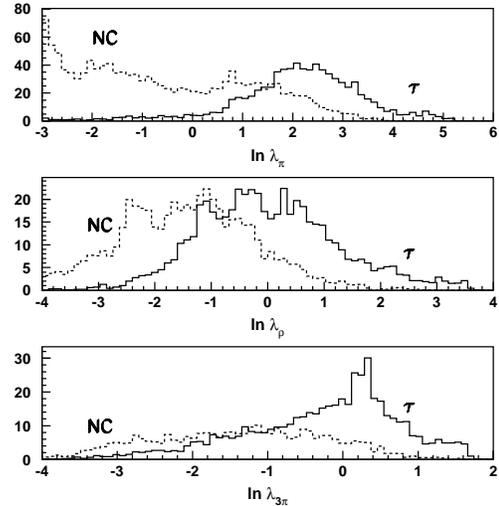,width=7.5cm}}
\caption{Likelihood ratio distributions, arbitrarily normalized, for
signal and background
events. The distributions are computed separately for each of the
considered hadronic channels: $\pi$, $\rho$, 3 prongs.}
\label{fig:ratios}
\end{figure}

The chosen variables are not independent one from another but show
correlations between them.
These correlations can be exploited to reduce the
background. In order to maximize the separation between signal 
and background, we use three dimensional likelihood functions 
${\cal L}(Q_T,E_{visible}, y_{bj})$ where
correlations are taken into account (see Figure~\ref{fig:ratios}). 

\begin{table}[tb]
\begin{tabular}{|c|c|c|c|c|c|c|}\hline
$\pi$ & $\rho$ & ${3\pi}$ & 
Top & Bot. & $P_\alpha$ & $P_\beta$  \\
Cut & Cut & Cut & Evts & Evts &  ($\%$) &  ($\%$) \\ \hline
0 & 0.5 & 0 & 112 & $134$ & 32.1 & $1.9$ 
($2.3\sigma$)\\
1.5 & 1.5 & 0 & 46 & $63$ & 24.8 & $0.7$ ($2.7\sigma$)\\
3 & $-1$ & 0 & 43 & $59 $ & 26.1 & $0.8$ 
($2.6\sigma$)\\
3 & 0.5 & 0 & 12 & {\boldmath $23$} & 
18.3 & {\boldmath $0.14$ $(3.3\sigma)$} \\ 
3 & 1.5 & 0 & 10 & $20$ & 18.8 & $0.16$ ($3.2\sigma$)\\
3 & 0.5 & $-1$ & 30 & $45$ & 21.9 & $0.4$ ($2.9\sigma$)\\
3 & 0.5 & 1 & 9 & $17$ & 25.9 & $0.5$ ($2.8\sigma$)\\ \hline

\end{tabular}
\caption{Expected background and signal events for different
combinations of the $\pi$, $\rho$ and $3\pi$ analyses. The considered
statistical sample corresponds to an exposure of 50
kton$\times$year. }
\label{tab:combi}
\end{table}

Table \ref{tab:combi} illustrates the 
statistical significance achieved by several selected combinations of the 
likelihood ratios for an exposure equivalent to 50 kton$\times$year. 
We take as the best combination the one with the lowest $P_\alpha$. 
This is achieved for the 
following set of cuts: $\ln \lambda_\pi > 3$, 
$\ln \lambda_\rho > 0.5$ and $\ln \lambda_{3\pi} > 0$. 
The expected number of NC background events amounts to 12 (top) 
while 12+11 = 23 (bottom) are expected. This corresponds to 
a $P_\alpha$ of 18.3$\%$. In the
case we consider $E_{visible}$ as the unique discriminating variable, a
similar number of background events is obtained demanding $E_{visible} 
> 14$ GeV. With this cut, the expected number of $\tau$ events 
is 7 and the $P_\alpha$ is 37$\%$. 
Therefore, for the same level of background, 
the approach using the ratio of three dimensional  
likelihood functions enhances the number of expected signal events by
approximately 50$\%$. 

\begin{figure}
\mbox{\epsfig{file=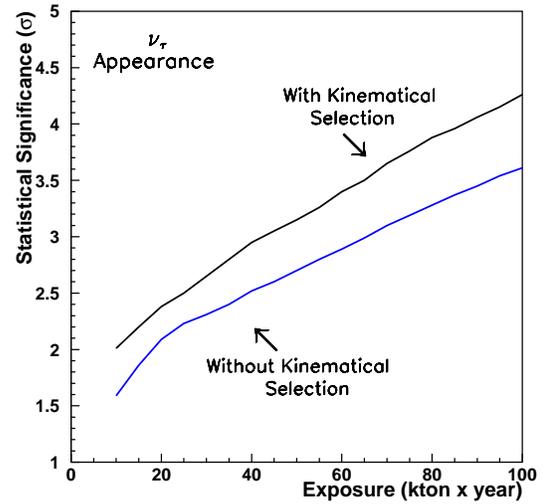,width=7.5cm}}
\caption{Probability for the measured excess of upward going events to
be due to a statistical fluctuation of the data as a function of the exposure.}
\label{fig:poisson}
\end{figure}

Finally, in figure \ref{fig:poisson} we present the Poisson probability 
$P_\beta$ for the measured excess of upward going events to be due to a
statistical fluctuation as a function of the exposure. The bottom curve 
corresponds to the case where no kinematical selection has been 
applied and only a cut on $E_{visible} > 6$ GeV is used. We see that 
for exposures around 30 kton$\times$year, in case 
we use the kinematical selection algorithm, the observed excess corresponds 
to a 2.6$\sigma$ effect. This effect is larger than $3\sigma$ for an 
exposure of 50 kton$\times$year.

\section*{Acknowledgments}
I thank the organizers of the NNN99 workshop, in particular,
C.K.~Jung. The help of A.~Bueno, M.~Campanelli, A.~Ferrari 
and J.~Rico is greatly acknowledged.


\end{document}